\documentclass[preprint,showpacs]{revtex4-1}
\newcommand{\be}{\begin{equation} }
\newcommand{\ee}{\end{equation}}

\usepackage{graphicx}
\usepackage{verbatim}
\begin{document}

\title{A terrestrial search for dark contents of the vacuum, such as dark energy, using atom interferometry}

\author{Ronald J. Adler}
\email[]{adler@relgyro.stanford.edu}
\affiliation{Gravity Probe B Mission, Hansen Experimental Physics Laboratory, Stanford University, Stanford CA 94309, USA\\and\\San Francisco State University, San Francisco, California 94132,USA}

\author{Holger Mueller}
\email[]{hm@berkeley.edu}
\affiliation{University of California, Berkeley, California 94720, USA}

\author{Martin L. Perl}
\email{martin@slac.stanford.edu}
\affiliation{Kavli Institute for Particle Astrophysics and Cosmology \\and  \\SLAC National Accelerator Laboratory, Stanford University, Menlo Park, California 94025, USA}

\date{\today \\  To To be submitted to Physical Review D}

\begin{abstract}

We describe the theory and  first experimental work on our concept for searching on earth for the presence of dark content of the vacuum (DCV) using atom interferometry. Specifically, we have in mind any DCV that has not yet been detected on a laboratory scale, but might manifest itself as dark energy on the cosmological scale. The experimental method uses two atom interferometers to cancel the effect of earth's gravity and diverse noise sources. It depends upon two assumptions: first, that the DCV possesses some space inhomogeneity in density, and second that it exerts a sufficiently strong non-gravitational force on matter. The motion of the apparatus through the DCV should then lead to an irregular variation in the detected matter-wave phase shift. We discuss the nature of this signal and note the problem of distinguishing it from instrumental noise. We also discuss the relation of our experiment to what might be learned by studying the noise in gravitational wave detectors such as LIGO.The paper concludes with a projection that a future search of this nature might be carried out using an atom interferometer in an orbiting satellite. The apparatus is now being constructed.
\end{abstract}


\maketitle

\section{INTRODUCTION}

\subsection{General remarks}

The nature of dark energy has been a dominant question in both cosmology and fundamental physics for the past decade \cite{Intro}. Our knowledge of the existence and properties of dark energy is entirely based on cosmological scale observations. This paper is directed towards experiments that might be able to measure effects of dark energy or any dark contents of the vacuum (DCV) on the laboratory scale. More precisely, the idea is to detect on the laboratory scale any hitherto unknown DCV that could give rise to behavior like that of the dark energy or cosmological constant on the cosmological scale. As our preferred technique, we describe atom interferometry experiments to detect DCV; we also describe the assumptions we make about the DCV. We believe this is the first suggestion for such experiments \cite{Perl1,Perl2}.

Recent work on atomic interferometry \cite{atom} has achieved high precision \cite{Mueller1}. For example, the gravitational acceleration $g$ at the earth's surface has been measured with a precision of about $10^{-9}$ \cite{peters}; this measurement has also confirmed the gravitational redshift with a precision of $7\times 10^{-9}$ \cite{Mueller1,redshifttheory}. Accuracies of $10^{-15}$ are expected in future atom interferometry laboratory experiments on the weak equivalence principle \cite{Kasevich1,Comp}. This precision is achieved through the small de Broglie wavelength of slowly moving atoms: the phase difference between the interferometers in past experiments can be millions of radians, whereas microradian changes are measurable. Moreover, as the beam splitters in atom interferometers are standing waves of laser light, the area enclosed by the interferometers' arms is given with the precision of a laser wavelength. Moreover, atoms have few and well characterized internal degrees of freedom, and couple in a well controlled way to their environment. When the great precision of atomic beam interferometry is combined with improvements that have recently been demonstrated such as large momentum transfer or common-mode rejection between conjugate interferometers \cite{SCI,BBB,BraggInterferometry}, new applications will become possible. Furthermore, the possibility of future operation in the nearly gravity-free environment of a spacecraft promises truly impressive possibilities \cite{redshifttheory,MWXG,AGIS}. 

\subsection{Organization of the paper}

There are four parts to this paper. Part 1, Sec. I and II, summarize the atom interferometry theory and describes our terrestrial experiment. Part 2, Sec. III and IV, discuss the severe experimental and observational limitations on our present knowledge of the nature of dark energy and the experimental philosophy behind this experiment. Part 3, Sec. V, VI, and VII, evaluate the reach of the experiment, and compare our experiments to other possibly relevant experiments. In these sections we are reluctant to speculate using complicated dark energy models, therefore we use pedestrian and limited models of how our experiment might detect dark energy density. Part 4, Sec. VIII, and IX discuss variations on our experimental design, and the advantage of moving from a terrestrial experiment to a space experiment.

For those who wish to scan the paper we recommend reading Sec. I, II, III and VIII.

\begin{figure}
\includegraphics[width =2in]{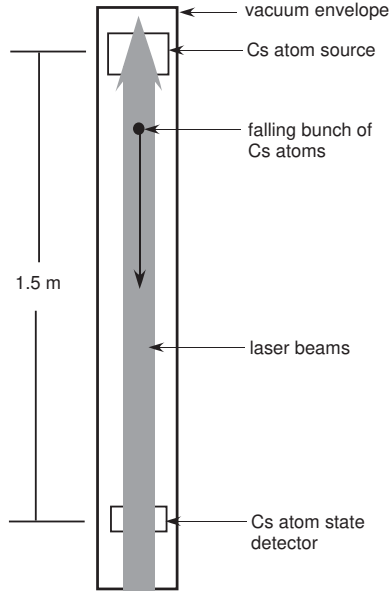}
\caption{A simplified schematic of a single atom interferometer in which a bunch of atoms is dropped vertically downward from the source at the top of the apparatus.}
\end{figure}

\begin{figure}
\includegraphics[width =3in]{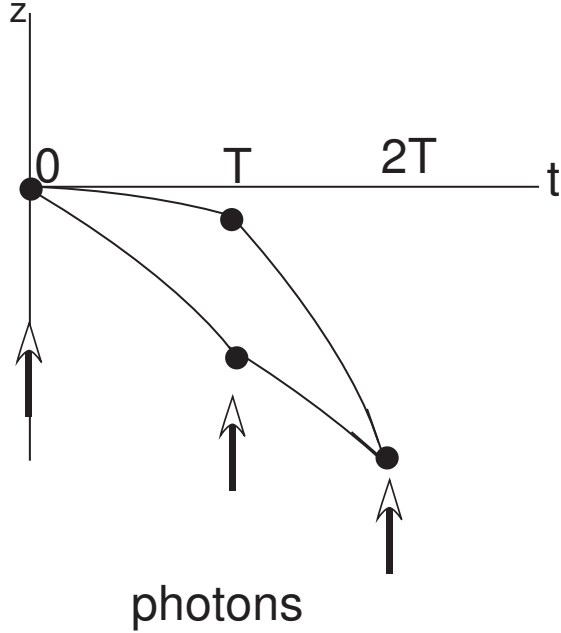}
\caption{Classical trajectories of atoms in  the drop experiment.}
\end{figure}
\subsection{Summary of atom interferometry theory}

Light-pulse atom interferometers use light to split an atomic matter wave, send it along two separate spacetime paths, and interfere the partial waves when the paths merge. The probability $P$ of detecting the atom at one output of the interferometer is given by the difference $\Delta\phi=\Delta \phi_1-\Delta \phi_2$ of the phases accumulated by the matter wave on the two paths by
\begin{equation}
P=[(1+\cos(\Delta \phi))]/2 .
\end{equation}
To further illustrate the interference mechanism, Fig. 1 shows a simplified schematic of a single atom interferometer in which a sample of atoms is dropped vertically downward from the source at the top of the apparatus \cite{directdrop}. An atom leaving the source is placed into a superposition of  two momentum eigenstates by interaction with the photons of counterpropagating laser beams; the first component of the matter wave receives zero momentum transfer while the second is given a downward momentum kick of $\Delta p$. The two matter wave packets then fall freely until a time $T$, whereupon they are given opposite $\Delta p$ momentum kicks. The two wave packets continue to fall until time $2T$ whereupon the first is given an upward kick of $\Delta p$ and the beams are recombined into a modulated single wave packet. The final beam population is measured by laser-driven fluorescence. Fig. 2 shows the classical paths of the beams to clarify the operation.

To calculate $\Delta\phi_{1,2}$, Feynman's path integral \cite{Feynman} or Bord\'e's 5D atom optics \cite{Borde5D} approach suggest themselves. Briefly, the atom is represented by a matter wave proportional to $\exp[(i/\hbar) \int L(x,\dot x) dt]$, where $\hbar$ is the reduced Planck constant and $L=\frac 12 m\dot x^2-V(x)$ is the Lagrangian;  $m$ is the particle mass, $x$ and $\dot x$ represent the coordinates and their time derivatives, and $V$ the potential energy. Integrating over all possible trajectories gives the propagator for a particle between two points. In the semiclassical limit, the path integral is dominated by the classical path. In this case, the phase  is given by an ordinary integral over the classical path and
\be
\Delta \phi_{1,2}=\phi_{{\rm a},1,2}+\phi_{{\rm p},1,2},
\ee
where
\be
\phi_{\rm a,1,2}=\frac{S_{\rm Cl,1,2}}{\hbar},
\ee
Here $\phi_{\rm a}$ denotes the phase shift due to the action and $S_{\rm Cl,1,2}=\int_{\rm 1,2} L(x,\dot x)dt$ is the classical action, evaluated along the interferometer arms 1 or 2. An additional influence $\phi_{\rm p,1,2}$ arises because whenever a photon is absorbed (or emitted), its phase is added to (or subtracted from) the phase of the matter wave. The final equations take this into account.

For a Mach-Zehnder interferometer in the earth's gravitational field with vertical laser beams, $V=mgz$, where $g$ is the acceleration of free fall, the leading order phase difference between the paths is given by
\be
\phi_{\rm MZ}=n k_{\rm{opt}} g T^2,
\ee
where $n$ is the number of photon momenta transferred to the atom in each beam splitter, $k_{\rm{opt}}$ is the laser wave number, and $g$ is the local gravitational acceleration in the laboratory. Thus, the sensitivity of the interferometer depends quadratically on the time of fall \cite{fountain1}.

\section{REALIZATION OF THE EXPERIMENT} 
\subsection{Basic design of the experiment}

\begin{figure}
\includegraphics[width =3in]{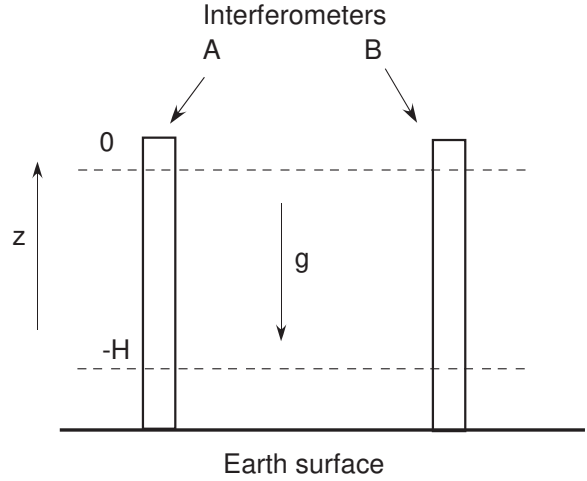}
\caption{Schematic illustration of the two interferometer system}
\end{figure}

Our experiment uses two interferometers, as close to identical in construction as is practical but separated in space. Fig. 3 shows a schematic. The system is designed to eliminate the effects of gravity and many sources of noise. For explanatory simplicity we take $g$ to be constant in space and time. The interferometers A and B produce phase shifts $\Delta \phi_A$ and $\Delta \phi_B$, which may be calculated as above, and give us a difference,
\begin{equation}
\Delta\Phi =\Delta\phi_{A}-\Delta\phi_{B}.
\end{equation}
If A and B are identical and no other interaction is present then clearly $\Delta\Phi=0$. The essence of our experiment is to null the phase shift difference $\Delta\Phi$ due to gravity with $\emph{very high precision, so that any other interaction becomes detectable}$.

Figure 4 shows a present embodiment of our experiment in which the two interferometers are enclosed in a single vacuum system and use a common laser system. 

\begin{figure}
\includegraphics[width =2in]{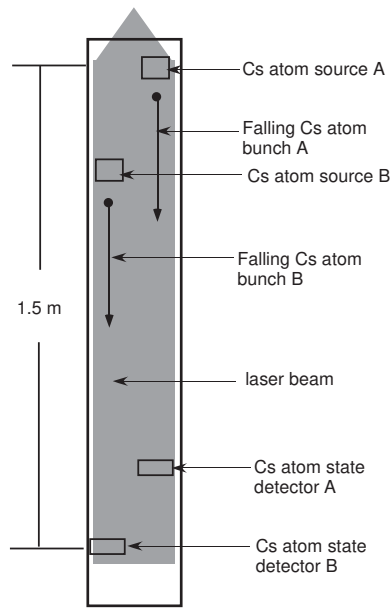}
\caption{The two interferometer system enclosed in a single vacuum envelope and using a common laser system}
\end{figure}

\subsection{Assumptions about dark energy inherent in this experiment}

The experiment depends on two fundamental assumptions about the DCV:
\begin{enumerate}
\item The DCV has a non-uniform spatial distribution, contrary to the cosmological constant model of dark energy. This leads to a potential dependent on the path of the falling atoms.
\item There is a non-gravitational force on the atoms due to the DCV, leading to the potential being larger than expected from gravity, Sec. VI.B.
\end{enumerate}
We know little about the DCV but it is cetainly not  stationary in the laboratory frame: It is  likely to exhibit temporal fluctuations as the Earth moves through the rest frame (if one exists) of DCV with a velocity comparable to typical galactic velocities of order several 100 km/s, in analogy to our velocity relative to the cosmic background radiation. Since an interferometer phase measurement requires about a second, $\Delta\Phi$  will be an average of an irregular signal. This is a crucial aspect of the experiment and will be discussed further in Sec. VI.C.

\section{OUR LIMITED KNOWLEDGE OF DARK ENERGY}

Dark energy first entered physics, in the guise of the cosmological constant, when Einstein introduced the cosmological term in the field equations of general relativity in 1917 in order to balance gravitational attraction and construct a static cosmological model \cite{Einstein}. The field equations including the cosmological term are
\begin{equation}
G_{\mu\nu} + \Lambda g_{\mu\nu} = (8\pi G/c^{4})T_{\mu\nu},
\end{equation}
where the cosmological constant is now almost universally denoted by $\Lambda$. The balance however was not stable, as soon noted by Eddington and others \cite{Eddington}. As is well-known, Einstein later abandoned the cosmological term as unnecessary, which became especially clear after Hubble's observation of the recession of galaxies and thus an expanding dynamic universe \cite{Hubble}. Others continued to discuss the cosmological term however and the idea was never forgotten \cite{Adler2}.

In 1998, measurements of the recession velocity of type Ia supernovae were made that implied the dynamic expansion of the universe is accelerating, and thus that the cosmological constant is nonzero and positive according to the standard cosmological interpretation \cite{Adler2,Riess}. This interpretation is based on the assumptions of homogeneity and isotropy in our region and throughout the universe. It is important to note that these assumptions and the relevant supernova physics are still being debated and tested, and the existence of the acceleration and of dark energy (or a cosmological constant) are still being questioned by some \cite{Demiaski}.

The cosmological term on the left side of the general relativity field equations, as originally used by Einstein, is consistent with and motivated by the mathematics of the Einstein tensor; if there were no energy-momentum source $ T_{\mu\nu} $ it would imply that spacetime on the cosmological scale is de Sitter space \cite{de Sitter}. But the cosmological term may be simply moved to the right side and interpreted as an energy momentum tensor source of gravity,
\begin{equation}
T_{\mu\nu}({\rm vac}) = -(\Lambda c^{4}/ 8\pi G)g_{\mu\nu}.
\end{equation}
The cosmological term then corresponds to a fluid with the rather unusual equation of state $p=-\rho$  , for which both pressure $p$ and density $\rho=\Lambda c^{4}/(8\pi G)$ are uniform in space and constant in time; this is the simplest conceptual version of dark energy \cite{Intro,Zeldovich}. In particular, there is no pressure gradient so the fluid does not exert a pressure force in the usual Newtonian sense. The present "standard model" or "concordance model" of cosmology includes this view of dark energy, with the equation of state parameter $w=p/\rho$  treated as an important observable quantity. The concordance model, including the value of $w$, is being vigorously tested using different types of astronomical observation \cite{dark energy}.

Dark energy can also be viewed as the ground energy state of the vacuum. This is qualitatively consistent with the idea that quantum fields (such as the electromagnetic field) have nonzero ground state energy that is constant in spacetime. However, quantum field theory suggests a spectacularly erroneous size for the cosmological constant, about $10^{120}$ times larger than indicated by observation, and other current theories do not seriously improve on this failure \cite{dark energy}. Thus it is, in our opinion, fair to say that present theory gives us no fundamental understanding of the nature of dark energy or of the cosmological constant \cite{Volovik}.

The contents of the present universe, according to the concordance model and observation, are about 75 \% dark energy, 20 \% cold dark matter, and 5 \% ordinary baryonic matter \cite{dark energy}. The total  density is consistent with a spatially flat Friedmann-Robertson-Walker universe in standard general relativity theory \cite{Steinhardt}. Note that the present era of accelerated expansion is qualitatively similar to the early inflation era (preceding the era of radiation), but the energy density is vastly smaller.

There are many speculations about the physical nature of dark energy beyond the simple cosmological constant view, far too many to consider here. One of the best known is a scalar quintessence field, $\psi$, which can cause acceleration much as the inflaton field in early universe inflation theory. Such a field can act much like a perfect fluid as a source of gravity with an effective equation of state parameter $w$ given approximately by
\begin{equation}
w=\frac{{\dot{\psi}^2}/2 -V(\phi)}{{\dot{\psi}^2}/2+V(\phi)}
\end{equation}
where $V$ is a self-interaction potential and the field is assumed to be uniform in space. Thus if the field is changing slowly it behaves like dark energy and is consistent with the concordance model \cite{Intro,Sadjadi}.

There are other interesting conjectures about the nature of dark energy, one of which is known as a tracker field. It is somewhat peculiar that the dark energy and dark matter energy densities have a ratio of about three at the present time, since that ratio changes from a very small value to a very large value as the universe expands and the dark matter becomes diluted. Thus some theorists speculate that there is a common source of the two and the ratio has a deeper explanation. We will not discuss such ideas here, but refer the reader to reference \cite{Rakhi}.

Dark matter has been considered on non-cosmological scales since it is supposed to clump readily in galaxies and galactic clusters, and it should presumably be observable as particles in the laboratory. But conventional dark energy has so far been associated only with the cosmological scale. On smaller scales the cosmological constant has quite small effects; roughly speaking it corresponds to a Newtonian quadratic potential \cite{Adler2}
\begin{equation}
\varphi=-(\Lambda c^2  /6) r^2
\end{equation}
where r is the distance from an arbitrary ``center''of the universe. There is thus a linear repulsive force between the objects in the universe. It becomes non-negligible compared to standard Newtonian attraction only on a scale larger than galaxies.

The all-important point we wish to make here is that we are at present totally ignorant about the physical nature of dark energy on smaller scales than cosmological, either from observation or theory. No observations or experiments have been made (or even seriously proposed!) due to the expected weakness of the associated gravitational forces involved. But in view of our ignorance of the fundamental nature of dark energy this expectation should be tested experimentally.

For example, on the cosmological scale it appears from both observation and theory that the dark energy in a spatial volume must be uniform and proportional to the volume, which is consistent with the concordance model. This clearly cannot be extrapolated to small scales, anymore than the large scale homogeneity and isotropy of the universe can be extrapolated to small scales, since that extrapolation would imply the nonexistence of galaxies and stars and physicists. We remain free to speculate on the small-scale nature of dark energy and to search for its properties, largely unconstrained by existing theory or observation. Thus there is no necessity to assume that its local density is uniform or time independent or comparable to its cosmological value, which is about 1\,GeV/m$^3$ \cite{Steinhardt}. Moreover, it is not necessary to assume that it interacts only gravitationally with ordinary matter, only that the interaction is sufficiently weak to have escaped detection so far. The only reasonable constraints are that, whatever its small scale nature, the average behavior of dark energy on the cosmological scale is similar to that of the cosmological constant.

To emphasize our ignorance of the small-scale nature of dark energy we point out an obvious analogy: on the cosmological scale both dark matter and baryonic matter are approximated as zero temperature uniform fluids whose non-gravitational interactions are irrelevant to cosmology; however on smaller scales they behave entirely differently, clumping to form complex structures and interacting much more strongly, so any extrapolation downwards would be total nonsense.

We close this section by noting that of the many alternative approaches to the cosmological constant problem there are some that do not involve viewing the dark energy as a physical fluid. In particular in one approach which we call ``higher order general relativity" or HGR, the field equations of general relativity are recast as third order differential equations for the metric tensor rather than the second order equations of standard general relativity theory; the third order equations have a close similarity to the equations of classical electromagnetism \cite{Cook}. For the cosmological problem the equations can be integrated once to obtain second order equations in which a constant of integration appears, which plays exactly the same role as the cosmological constant.  Thus, the cosmological constant may be re-interpreted as a purely mathematical constant of integration with an arbitrary value, unrelated to a physical energy density. One peculiar property of the theory is that radiation also appears in a very analogous way! It is presently not clear if HGR has physical content beyond providing a different view of the cosmological constant as a constant of integration.

\section{PHILOSOPHY OF OUR EXPERIMENT}

\subsection{Philosophy}

Although it is less than two decades since the astonishing discovery of the dark energy phenomenon, the experimenter or observer desirous of  investigating dark energy already faces difficulties. Astronomical observations of the dark energy phenomenon are becoming increasingly precise. But, as discussed in the previous section these observations are on the cosmological scale and are unlikely to teach us anything of the essence of dark energy, is it an energy field, is it related to matter, is it just a term in the equations that describe general relativity, is it a fluid? Or is it some other phenomenon that we have never thought about?  During the doctoral research of one of us (M.L.P.) his supervisor, the laureate Isadore Rabi, repeatedly reminded him that "Physics is an experimental science". This is the spirit behind this experiment.

We are looking for a new way to microscopically penetrate the mystery of dark energy and more generally to penetrate the puzzles associated with the vacuum. Atom interferometry provides such a way for the following reasons:
\begin{itemize}
\item This technology depends on the simple , well understood interaction of a single atom with a photon.
\item This technology is becoming increasingly precise.
\item The cost of the components is deceasing as industrial use of lasers and electro-optics increases.
\item The terrestrial version of this experiment can be constructed and carried out by a few people thus providing the ability to rapidly improve the experiment.
\item The terrestrial version of this experiment is inexpensive compared to most modern astronomical instruments intended to study dark energy.
\item The terrestrial experiment can lead to the experiment being carried in earth orbit on a space platform.
\end{itemize}

\subsection{Comparison of dark energy density with electric field density} 

Another view of our philosophy is made evident by our early comparison  between dark energy and the electric field \cite{Perl1,Perl2}.

Consider a weak electric field $E$ = 1 V/m. Using
\begin{equation}
\rho_{ef} = \epsilon_{0}E^{2}/2,\quad
\rho_{ef}=4.4\times10^{-12} \mbox{  J/m$^3$}
\end{equation}
Hence the energy density of this electric field is 100 times smaller than the dark energy density, $\rho_{DE}$ = $6.3\times10^{-10}$  J/m$^{3}$, yet this weak electric field is easily detected and measured. This realization first started our thinking about the possibility of direct detection of dark energy. The point is that dark energy density is small but it is not zero. And to the experimenter anything that is not zero may be, and should be, a subject for detection and measurement.

Of course, it is easy to sense and measure tiny electromagnetic fields due to the relatively strong coupling; on the other hand there are obviously severe experimental problems in detecting dark energy or DCV density.
\begin{itemize}
\item{Unlike an electric field in the laboratory, we cannot turn dark energy on and off.}
\item{We do not expect there is a zero dark energy field  that could be used as an experimental reference. }
\item{Even if the dark energy density should have a gradient, we do not know what force  it exerts on a material object as discussed in Sec. IV. }
\end{itemize}
The point is that while these are severe experimental problems, the average dark energy density is not zero, and given the correct experimental method, variations in it may be detected.

\section {PARAMETERIZATION OF THE UNKNOWN CONTENTS OF THE \\ VACUUM}

\subsection{Basic physical concept}

In accord with the comments in the previous section we will consider the possibility that the DCV are not uniform in space and may involve non-gravitational interactions. We will describe an ad hoc parametric model of the DCV on a small scale that is consistent with the properties of the dark energy on the cosmological scale. It allows us to calculate phase shifts for rather general interactions in terms of a few parameters. Of necessity the model is very speculative but our hope is that it captures in a phenomenological way some of the properties that the DCV may have.

In this experiment the interferometers are fixed to the earth. The earth is spinning and moving in the Galaxy and the Galaxy is moving in the CMB frame with a velocity about 300 km/s. In any case, we do not expect the DCV to be tied to the earth. Therefore the sought signal will average over many samplings of the DCV and will \emph{be an irregular signal}.

Recall that the cosmological term of the general relativity field equations corresponds to an effective Newtonian repulsive quadratic potential energy function  Eq. (9) or a potential energy.
\begin{equation}
U=m\varphi=-m\Lambda c^2  r^2/6
\end{equation}
This appears to represent quite well the effective behavior of the dark energy on the cosmological scale; on a smaller scale we simply do not know anything about its behavior.  Thus we will reasonably assume it behaves spatially in an irregular fashion. The rough magnitude of the variations in the potential energy we denote bt   $\tilde{U}_D$ (which may locally far exceed the average of the potential), and the spatial extent of the variation by  $R$. This view is qualitatively equivalent to assuming that the DCV occurs in lumps randomly in space; thus one atomic beam in a double interferometer may pass through a lump while the other beam does not. Alternatively the variations in the DCV   \emph{may of course have a regular structure.}

\subsection{Quantitative phase shift calculation}

For a quantitative description, we assume $U(X)$ depends on position $X$ but is time-independent in some rest frame. While it is unknown what property of the test particle the nongravitational coupling of the DCV couples to (e.g. baryon number etc), it is to be expected that any such coupling is roughly proportional to the mass $m$ of the particle; we thus take it to be $mu'$, where $u'$ describes the local value of the fluctuating potential. To model the stochastic fluctuations, we use a superposition of plane waves
\be
U(X)={\rm Re}\left(\int_0^\infty mu'(k) e^{ikX}dk\right),
\ee
where $U$ is the potential energy, $mu'(k)$ is the spectral density of the potential generated by the DCV, and Re denotes the real part. In the following, we will drop the Re from the equations; taking the real part is implied. This model is consistent with cosmology, because the fluctuations average out over large distances for a suitable low-frequency cutoff of the $u'(k)$.

The Earth moves through the rest frame with velocity $v,$ so that $X=x-vt$ and
\be
U(x)=\int_0^\infty mu'(k) e^{ikx-i \omega_k t}dk,
\ee
where $\omega_k=kv$. We will now restrict attention to one Fourier component $u(k)$. If we assume that the dimensions of the apparatus are much smaller than $1/k,$ we can approximate
\be\label{pot}
U(x,t)=mu(k)e^{(ikx-i\omega_kt)} \approx mu(k) e^{-i \omega_k t} [1 + ik (x-x_0) - \frac 12 k^2 (x-x_0)^2],
\ee
where $x_0$ is the lowest point of the interferometer. The constant term produces no effect in the output signal, so we ignore it from here on.

To calculate the phase shift $\phi$ between the arms of one interferometer, we use a perturbative approach in which the phase shift is calculated by integrating the perturbing U potential over the unperturbed paths of the Mach-Zehnder configuration shown in Fig. 2:
\be
x_1(t)=\left\{\begin{array}{cc} (v_L+v_r)t-\frac12 gt^2 & t<T,\\
v_rT+v_Lt-\frac12 gt^2 & t>T\end{array}\right.
\ee
and
\be
x_2(t)=\left\{\begin{array}{cc} v_Lt-\frac12 gt^2 & t<T,\\
v_r(t-T)+v_Lt-\frac12 gt^2 & t>T\end{array}\right.
\ee
Here, $v_L$ is the initial (launch) velocity of the atom and $v_r$ the recoil velocity {\em i.e., the velocity difference between  atoms in the two interferometer arms}. The phase between the interferometer arms is given by
\be
\phi=\frac 1\hbar \left(\int_{\rm Path\,\,1} U(x,t)dt-\int_{\rm Path\,\,2} U(x,t)dt\right).
\ee
The phase {\em difference} between the two interferometers separated by $L$ is given by
\be
\Delta \Phi=-4\left(\frac{L}{v}\right) \left(\frac{v_r}{v}\right) \left(\frac{m u(k)}{\hbar}\right) \sin^2\left(\frac{k v T}{2}\right) \cos\omega_k t_0.
\ee
Here, $t_0$ is the time at which the central interferometer pulse is applied. If we substitute $v_r=n \hbar k_{\rm opt}/m$, where $n$ is the number of photon momenta transferred to the atom per beam splitter, and $k_{\rm opt}$ the laser wavenumber,
\be
\Delta \Phi=-4\left(\frac{L}{v}\right) \left(\frac{nk_{\rm opt}}{v}\right) u(k)\sin^2\left(\frac{k v T}{2}\right) \cos\omega_k t_0.
\ee
To get some intuition for the significance of $u$, note that $u(k) k\equiv a(k)$ is the acceleration at $x_0$ of the test particle due to the DCV, hence
\be
\Delta \Phi=-4\frac{L nk_{\rm opt}a(k)}{v\omega_k}\sin^2\left(\frac{\omega_k T}{2}\right) \cos\omega_k t_0.
\ee
For a numerical estimate, we use $v=300$\,km/s, $k_{\rm opt}=10^7$/m and obtain
\be
\Delta \Phi=-21  n\left(\frac{L}{{\rm m}}\right)\left(\frac{a(k)}{{\rm m/s}^2}\right)\left(\frac{\omega_k}{2\pi 1/{\rm s}}\right)^{-1} \sin^2\left(\frac{\omega_k T}{2}\right) \cos\omega_k t_0 {\rm rad}.
\ee
This has to be compared with the resolution of the instrument, which may be on the order of microradians to milliradians. For an estimate, we assume that the average atom flux through the interferometer is $\eta$. We also assume that the interferometer resolution is shot-noise limited. This leads to a phase noise spectral density of $\sqrt{\eta}$. If the interferometer can be run for a long time, the resolution towards coherent signals will improve as the square-root of the integration time. By choosing the pulse separation time $T$, the sensitivity towards a given frequency $\omega_k$ can be maximized; this maximum decreases as $1/\omega_k$.

\section{Comparison with other experiments}

\subsection{Comparison with gravitational wave detectors}

In this comparison we use the dark energy density model of the previous section in which the density is described by a spatial Fourier series. A gravitational wave detector, such as the Laser Interferometer Gravitational Wave Observatory LIGO \cite{LIGO1}, is modeled here as a Michelson interferometer consisting of mirrors which are freely floating at the frequency scales of interest. These mirrors will then accelerate according to Newton's  second law, where the force is given by Eq. (\ref{pot}) as
\be
F(x,t)=\frac{\partial U}{\partial x} = ik mu(k)e^{ikx-i\omega_kt}.
\ee
Note that we have not assumed that the arm  length of the gravitational wave detector is short relative to the length scales of the DCV fluctuations; we do, however, assume $v\ll c$. We integrate Eq. (22) twice to obtain the position modulation of the mirrors,
\be
\delta x= -\frac{ik u(k)}{\omega^2} e^{ikx-i \omega_k t}.
\ee 
The difference between the modulations of two end mirrors that are assumed to be separated by $L$ is given by subtracting the above for $x=0$ and $x=L$. We divide the result by $L$ to obtain $h$, the strain signal:
\be
h= -\frac{ia(k)}{\omega_k^2L}e^{-i\omega_kt}\left(1-e^{ikL}\right).
\ee 
This can be compared to the published noise equivalent strain magnitudes of existing gravitational wave detectors.

\begin{figure}
\centering
\includegraphics[width=4in]{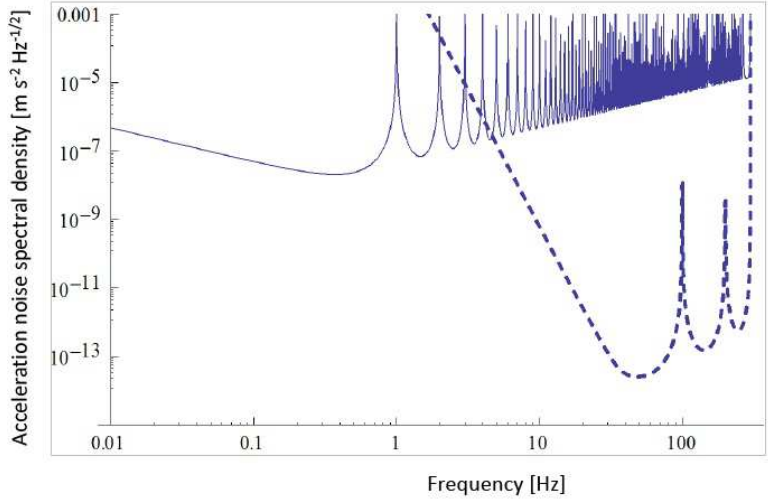}
\caption{\label{phase3a.eps} A comparison of the equivalent acceleration noise spectral densities of the proposed atom interferometer DCV detector (solid line) and LIGO as used as a DCV detector (dashed line).  Equivalent acceleration noise spectral density in m/s$^2/\sqrt{\rm Hz}$ attainable with a dual atom interferometer assuming $L=10$\,m, $k_{\rm{opt}}=10^7$/m, $n=10$, $v=300$\,km/s, $T=1\,$s and atom flux of $\eta=10^{8}$/s.}
\end{figure}

Fig. \ref{phase3a.eps} shows a comparison of the equivalent acceleration noise spectral densities of the proposed atom interferometer DCV detector and LIGO as used as a DCV detector \cite{LIGO2}. The graphs are plotted in terms of $a$ per square root Hertz; any $a$ that lies above the curves by a sufficient signal to noise margin can potentially be detected. 

It is obvious that LIGO is much more sensitive for frequencies above 3\,Hz; at the assumed velocity, this corresponds to DCV potential fluctuations having a wavelength of 100\,km or less.  The proposed atom interferometer, however, is better suited for the low frequency, long-wavelength fluctuations below 3\,Hz or above 100\,km.

\subsection{Consideration of  the expected gravitational force of dark energy}

It is  important to ask if a potential $\tilde{U_D}$ consistent with a gravitational interaction would be detectable. If the interaction were gravitational we would expect on dimensional grounds that very roughly
\begin{equation}
\tilde{U_D} \approx \frac{GmM_D}{R} \approx \frac{Gm(\rho_D/c^2)R^3}{R} = Gm\rho_D R^2/c^2,
\end{equation}
where $R$ is the blob size, $M_D$  is the mass of the blob and $\rho_D$ is the energy density of the blob. Thus from Eq. (17)
\begin{equation}
\Delta \phi \approx  \frac{\tilde{U_D R}}{\hbar v} \approx \frac{Gm\rho_D R^3}{\hbar v c^2}.
\end{equation}
Using the dark energy density to illustrate, $\rho_D \geq \rho_{DE}$
$\approx 10^{-9} $ J/m $^3$, we obtain, with $m\approx 10^{-25}$ kg and $R \approx 1$ \,m
\begin{equation}
\Delta\phi \approx 10^{-32} \mbox{\,rad}.
\end{equation}
A phase shift of such an extremely small magnitude is not observable with an atom interferometer, hence it is clear that we must assume a non-gravitational interaction in this experiment.

\subsection{Other possibly relevant experiments}

There have been numerous recent and elegant measurements  of $g$ and $G$; and precise  tests of the validity of general relativity and of the equivalence principle. An example of the latter is the  test of Schlamminger {\em et al.}  \cite{Schlamminger} using a rotating torsion balance to compare beryllium and titanium. They found the upper limit on a deviation from the equivalence of inertial mass and gravitational mass to be $10^{-15}$ \,m/s$^2$. This test is only relevant to the detection of dark energy density if one assumes dark energy acts differently on  beryllium and titanium. 

\section{SENSITIVITY OF THE EXPERIMENT}

An example of present experimental sensitivities is given by  the measurement of $g$ in an atomic fountain interferometer  as described by Chu \cite{Chu} and by Chung and his coworkers .\cite{Chung} In a single interferometer,  $g$  has been measured with a precision of $10^{-9}g$ =$10^{-8}$ \, m/s$^2$. The sensitivity is limited by noise caused by mechanical vibrations, laser frequency fluctuations or optical component drifts and the like \cite{Gouet1,Gouet2,Merlet}. If there were no noise, the maximum  sensitivity would be set by the shot noise of the atomic beam.

We next calculate approximately this maximum sensitivity for our experiment. Consider dropping Cs atoms from a height of  $H$ = 1\, m. 
The total phase shift in the .45\, s fall is given by
\begin{equation}
\Delta\phi=k_{eff}g T^{2} =6.51\times 10^{7}\, \mbox{rad}.
\end{equation}
Here, $k_{eff}$ is the total momentum change in terms of wave number when the Cs atom absorbs two photons of wave length 852\,nm.

Define $a=g/\Delta\phi$ in units of  $m/s^2$. In this example $a=1.51\times 10^{-7} \,m/s^2$. The shot noise error in the measurement of $g$ is given by the formula
\begin{equation}
\sigma_g = a/\sqrt{(N(T/.45\,s))}.
\end{equation}
Here $N$ is the number of Cs atoms in a single drop and $T$ is the time in seconds. This equation contains the ubiquitous square root of the total number of counts in a time $T$. We evaluate this last equation for various relevant values of $N$ and $T$.

For a moderately precise interferometer using $N=10^{6}$. 
\begin{equation}
\sigma_g(10^{6},  .45\, \mbox{s})=1.5\times 10^{-10}\, \mbox{m/$s^2$}.
\end{equation}
The most precise interferometers reach up to $N=10^{8}$ and for various times we find
\begin{eqnarray}
\sigma_g(10^{8},  10^2\, \mbox{s})=1.0\times 10^{-12} \, \mbox{m/$s^2$},\\
\sigma_g(10^{8},  10^4\, \mbox{s})=1.0\times 10^{-13}\,  \mbox{m/$s^2$},\\
\sigma_g(10^{8},  10^7\, \mbox{s})=3.2\times 10^{-15}\,  \mbox{m/$s^2$}.
\end{eqnarray}
Here the $10^7$\, s example is representative of the long data runs intended for this experiment.
If we can increase $N$ to $10^{10}$ the shot noise limit becomes
\begin{equation}
\sigma_g(10^{19},  10^7\, \mbox{s})=3.2\times 10^{-16} \, \mbox{m/$s^2$}.
\end{equation}

Since we are looking for an irregular signal in this experiment, it is obvious that we must substantially reduce the noise sources. Noise is a major concern in the community of experimenters using atom interferometry \cite{Gouet1,Gouet2,Merlet}. Our approach is pay close attention to mechanical vibrations and optical noise. For example, we intend to operate the entire double spectrometer from a single laser and the first realization of the experiment uses the compact design of Fig. 4.

\section{ OTHER GENERAL FORMS OF THE EXPERIMENT}

\subsection{The atomic fountain interferometer}

In the discussions in this paper we have emphasized the atomic drop interferometer for simplicity. The experiment can also be performed using the atomic fountain method \cite{Chu, Chung}. Indeed for the same height apparatus the atomic fountain method gives four times the sensitivity compared to the drop method.

\subsection{Spread double interferometers}

Figure 6  shows schematically two forms of the experiment in which the two interferometers are separated in distance either vertically or horizontally. These designs are more susceptible to uncancelled noise from mechanical vibrations and laser
signal transport instabilities. While we know nothing about the possible nature of inhomogenities in the dark energy density,  our intuition is that spread double interferometers provide a larger field of exploration.
\begin{figure}
\centering
\includegraphics[width=3in]{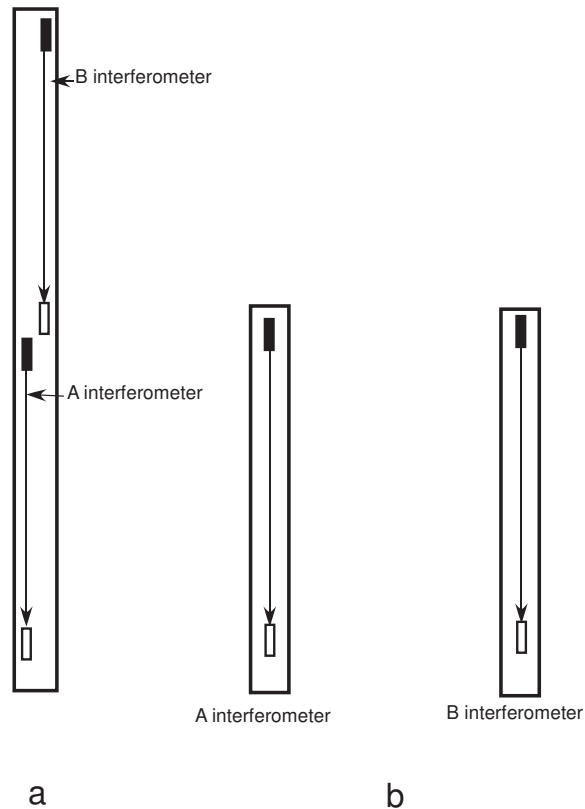}
\caption{ Schematic illustrations of spread double interferometers. (a) A vertically spread interferometer, the vertical distance would be of the order of several meters. (b) A horizontally spread interferometer, the horizontal distance between interferometers would be of the order of several meters.}
\end{figure}

\subsection{Use of a single interferometer}

Our experiment uses a pair of interferometers, but we have also considered the experiment using a single interferometer. Could one use a single interferometer,  obtaining an indication of DCV by analyzing the signal as a function of time, say at half hour intervals? However we do not see how to solve the problem of canceling  the noise  from mechanical vibrations and laser instabilities.

\section{USING THE METHOD ON AN EARTH ORBIT PLATFORM}

A great deal of design, experimental and theoretical work has been done on putting an atom interferometer into space using an earth orbit satellite although this has not yet been accomplished \cite{space}. Operating an atom interferometer in space leads to substantially improved measurements, both fundamental and geophysical \cite{space}.

It is obvious that space operation has two benefits for our search for DCV:

\begin{enumerate}
\item The problem of precise nulling of $g$ is eased, although gravitational tidal effects still exist.
\item The measurement period for an atom bunch can be substantially extended.
\end{enumerate}
On the other hand there are the well known problems and the substantial additional costs of space operation:

\begin{enumerate}
\item Long periods of uniform data acquisition are required compared to current, terrestrial atom  interferometer experiments that run for short time periods.
\item Long periods of perfect apparatus performance are required compared to current, terrestrial atom  interferometer experiments  that may require frequent adjustment or maintenance.
\item The apparatus must be space qualified.
\end{enumerate}

It will be a useful future task to make a quantitative study of carrying out our searches in space.

\begin{acknowledgments}

We are grateful to President John Hennessy of Stanford University for providing University funding and grateful to Director Persis Drell of the SLAC National Accelerator Laboratory for providing a laboratory for the experiment.  

We thank Dennis Ugolini for illuminating help on the comparison with LIGO.

We thank Helen Butler, Adrienne Higashi, David Jackson, David MacFarlane and Frank O'Neill of the SLAC National Accelerator Laboratory for their continual support. We thank Ziba Mahdavi of the Kavli Institute for Particle Astrophysics and Cosmology for her continual support

Ronald Adler is grateful to members of the  Stanford University GPB theory group, James Bjorken, and John Mester.

Martin Perl thanks Jason Hogan of Stanford University for his patient tutoring on the photonics needed for this experiment and he thanks Mark Kasevich for a critical discussion on this experiment, a discussion that led to a major design change.

\end{acknowledgments}

\appendix


\begin{thebibliography}{99}

\bibitem{Intro} W. I. Freedman and E. W. Kolb, Cosmology, chapter 1 of  The New Physics for the Twenty-First Century, edited by Gordon Fraser, (Cambridge University Press, Cambridge UK, 2006) R. J. Adler, Gravity, chapter 2, {\em ibid}. For a comprehensive discussion of the history see the online encyclopedia article by A. Riess (primary contributor): www.brittanica.com/Ebchecked/topic/1055698/dark-energy. A somewhat different view is given by E. Bianchi, C. Rovelli, and R. Kolb, Nature {\bf 466}, 321, (2010).

\bibitem{Perl1}  M. L. Perl and H. Mueller, arXiv 1001.4061v1 (2010).

\bibitem{Perl2} M. L. Perl,arXiv 1007.1622v1 (2010).

\bibitem{atom} There is a large literature on atom interferometry but no textbooks or monographs giving a uniform introduction to the subject. For practical introductions to laser cooling and atom interferometry. We recommend: C. Wieman, G. Flowers, and S. Gilbert, Am. J. Phys.  {\bf 63}, 317 (1995); A. S. Melish and A C Wilson, Am. J. Phys. {\bf 70}, 965 (2002); C. J. Foot, Cont. Phys. {\bf 36}, 369 (1991; O. Carnal and J. Mlynek, Phys. Rev. Lett.  {\bf 66}, 2689 (1991); A. Peters {\em et al.} , Phil. Trans. R. Soc. London a,{\bf 355}, 2223 (1997); M. de Angelis {\em et al.} , Meas. Sci. Technol. {\bf 20}, 022001 (2009); A. D. Cronin {\em et al.}, Rev. Mod. Phys. {\bf 81}, 1051 (2009), A. Peters {\em et al.} , Metrologia {\bf 38}, 25 (2001),R. M. Godun {\em et al.} , Contemp. Phys., {\bf 42}, 77 (2009),Q. Bodart {\em et al.} , App. Phys. Lett. {\bf 96}, 134101 (2010).

\bibitem{Mueller1} H. Mueller, A. Peters, and S. Chu, Nature {\bf 463}, 926 (2010).

\bibitem{peters} A. Peters, K.-Y. Chung, and S. Chu, Nature {\bf 400} 849852 (1999); S. Merlet {\em et al.} , Metrologia {\bf 47}, L9 (2010).


\bibitem{redshifttheory} M. A. Hohensee  {\em et al.}, arXiv:1009.2485 (2010).

\bibitem{Kasevich1} Private communication from M. Kasevich.

\bibitem{Comp} For a comprehensive discussion in the context of equivalence principle experiments see S. Dimopoulos {\em et al.} , Phys. Rev. Lett. {\bf 98}, 111102, (2007); S. Dimopoulos {\em et al.}, Phys. Rev. {\bf  D78}, 042003, (2008).

\bibitem{BraggInterferometry} H. Mueller {\em et al.}, Phys. Rev. Lett. {\bf 100}, 180405 (2008).

\bibitem{BBB} Holger Mueller {\em et al.},  Phys. Rev. Lett. {\bf 102}, 240403 (2009).

\bibitem{SCI} Sheng-wey Chiow {\em et al.},  Phys. Rev. Lett.  {\bf 103}, 050402 (2009).

\bibitem{MWXG} W. Ertmer {\em et al.},  Exp. Astron. {\bf 23}, 611-649 (2009).

\bibitem{AGIS} J. M. Hogan {\em et al.},  arXiv:1009.2702 (2010).

\bibitem{directdrop} Q. Bodart {\em et al.} , App. Phys. Lett. {\bf 96}, 134101 (2010).

\bibitem{Feynman} R. P. Feynman and A. R. Hibbs, Quantum Mechanics and Path Integrals (McGraw Hill, New York, 1965) ch.2.

\bibitem{Borde5D}Ch. J. Bord\'{e}, European Phys. J. - Special Topics, {\bf 163}, 315 (2008).

\bibitem{fountain1} Similar relations occur for a fountain atom interferometer; see Ref. \cite{Mueller1}.

\bibitem{Einstein} A. Einstein, Sitzber. Preuss. Akad. Wiss., 142, 1917.  English translation is in The Principle of Relativity (Methuen, 1923, reprinted by Dover Publications), p. 35.

\bibitem{Eddington}  A. S. Eddington, Monthly Notices of Roy. Astr. Soc. {\bf 90}, 668, (1930).

\bibitem{Hubble} E. P. Hubble, Astrophys. J. {\bf 84}, p 270, (1936).

\bibitem{Adler2}  R. J. Adler, M. Bazin, M. Schiffer, Introduction to General Relativity, 2nd. ed. (McGraw Hill, NY, 1975), sec. 13.2 and 12.5.

\bibitem{Riess} A. Riess, {\em et al.}, Astronomical Journal {\bf 116}, 1009 (1998); S. Perlmutter, {\em et al.}, Astrophys. J. {\bf 517} ,565 (1999).

\bibitem{Demiaski}    M. Demiaski and E. Piedipalumbo, in The Tenth Marcel Grossmann Meeting, on Recent Developments in Theoretical and Experimental General Relativity, Gravitation and Relativistic Field Theories, (World Scientific, Singapore, 2010) p. 1782; For evidence favoring the standard interpretation see P. Zhang and A. Stebbins, arXiv:1009.3967 (2010).

\bibitem{de Sitter}  W. de Sitter, Proc. Roy. Acad. Sci. (Amsterdam), {\bf 19} 1217 (1917); {\bf 20}, 229, (1917); {\bf 20}, 1309 (1917);  Mon. Not. Roy. Astron. Soc. {\bf 78}, 3 (1917).

\bibitem{Zeldovich} Y.B.Zel'dovich, JETP letters {\bf 6}, 316 (1967); Soviet Physics Uspekhi {\bf 11}, 381 (1968).

\bibitem{dark energy} A. Albrecht {\em et al.}, {\em Report of The Dark Energy Task Force}, arXiv:astro-ph/0609591v1(2006).

\bibitem{Volovik} For a contrary view concerning emergent gravity see G. E. Volovik,arXiv: 0604062v4 (2006).

\bibitem{Steinhardt} See overview notes by P. J. Steinhardt at www.physics.princeton.edu/-Steinh/steinhardt.pdf (2003).

\bibitem{Sadjadi} H. M. Sadjadi, Eur.Phys.J. {\bf C66}, 445 (2010).

\bibitem{Rakhi} R. Rakhi and K. Indulekha, arXiv:0910.5406 (2009).

\bibitem{Cook} R. Cook , arXiv:0810.4495v2 (2009).

\bibitem{LIGO1} P. R. Saulson, {\em Fundamentals of Interferometric Gravitational Wave Detectors} (World Scientific , Singapore, 1994).

\bibitem{LIGO2} We have used the LIGO noise information from D. Ugolini, Document no. LIGO-G100214 (2010).

\bibitem{Schlamminger}S. Schlamminger {\em et al.}, Phys. Rev. Lett. {\bf 100}, 041101(2009).

\bibitem{Chu} S. Chu, Rev. Mod. Phys. {\bf 70}, 685 (1998).

\bibitem{Chung} K-Y Chung {\em et al.}, Phys. Rev. {\bf D80}, 016002 (2009).

\bibitem{Gouet1} J. LeGou\"{e}t {\em et al.}, Appl. Phys.{\bf B92}, 133 (2008).

\bibitem{Gouet2}J. LeGou\"{e}t {\em et al.}, Eur. Phys. J,{\bf D44}, 419 (2007).

\bibitem{Merlet}S. Merlet, {\em Gravity, Geoid and Earth Observations} (Springer-Verlag, Berlin, 2010) ed. S. P. Mertikas.

\bibitem{space} O. Carraz {\em et al.}, App. Phys. {\bf B97}, 405, (2009), D. A. Binns {\em et al.}, Advances Space Sci. {\bf 43},1158 (2009), F. Sorrentino {\em et al.}, arXiv:1003.1481v1 (2010).

\end{thebibliography}
\end{document}